\documentclass[12pt,preprint]{aastex}
\begin{document}
\title{Evolution of the Pancaking Effect in a $\Lambda$CDM Cosmology}
\author{Yookyung Noh and Jounghun Lee}
\affil{Department of Physics and Astronomy, FPRD, Seoul National
University, Seoul 151-747, Korea} 
\email{ykyung@astro.snu.ac.kr,jounghun@astro.snu.ac.kr}

\begin{abstract}
We explore the evolution of the large-scale anisotropy in the velocity field 
caused by the gravitational pancaking effect assuming a $\Lambda$CDM universe. 
The Millennium Run halo catalogs at four different redshifts, $z=0,\ 0.5,\ 1$ 
and $z=2$ are analyzed to find that the pancaking effect starts to intervene 
the hierarchical structure formation at redshift $z=2$ when a characteristic 
pancake scale is around $3\ h^{-1}$Mpc. It is also clearly shown how the 
degree and scale of the pancaking effect changes with time.
An analytic model based on the Zel'dovich approximation is presented to 
explain quantitatively the evolution of the velocity-pancake alignment. 
A cosmological implication of our finding and a possibility of 
detecting a signal in real universe are discussed.
\end{abstract}

\section{INTRODUCTION}

The observed filamentary distribution of the large scale structure in the 
universe, which are often called the cosmic web, have so far motivated 
plenty of works. After the ingenious breakthrough made by \citet{bon-etal96}, 
a theoretical framework has been provided within which the characteristic 
properties of the cosmic web can be explained qualitatively in terms of 
the tidal influences in the universe dominated by the cold dark matter (CDM).

The influence of the tidal forces on the orientations and distributions 
of the large scale structure has been a target of many fruitful studies 
which either numerically or observationally proved that the tidal effect 
is strongest near the pancakes.

A pancake refers to the two-dimensional sheet-like structure in the universe, 
which form through gravitational collapse along the local tidal field. 
The first order Lagrangian perturbation theory, the Zel'dovich approximation 
\citep{zel70}, is probably the most popular model for the pancake formation 
which was originally suggested assuming the hot dark matter (HDM) dominated 
universe. The cosmic web theory, however, showed brilliantly that the 
presence of pancakes are also very possible in the CDM-dominated universe 
due to the large scale coherence of the tidal field. N-body simulations 
of CDM particles indeed demonstrated that the first collapsed objects look 
like pancakes \citep{sha-etal95}.

The large-scale coherence of the tidal field that is responsible for 
the formation of pancakes in turn causes the dark matter halos near or on 
the pancakes to have strong spatial correlations in the orientations
\citep{ara-etal07,bru-etal07,hah-etal07}.
The reason that the degree of the tidally induced alignments is highest
near the pancakes can be understood given the fact that the pancakes 
are the first collapsed objects, being still in linear regime while 
the filaments are more severely modified by the nonlinear process.

The formation of pancakes induces not only the spatial correlations in the 
halo orientations but also the anisotropy in the velocity field. 
In our previous paper \citep[][hereafter, NL06]{noh-lee06}, we have for the 
first time detected a $5\sigma$ significant signal of the velocity anisotropy  
induced by the pancaking effect in the high-resolution N-body simulation. 
In spite that the signal is found to be very weak because of the non-conserved 
nature of the velocity field, the velocity anisotropy holds a crucial key to 
understanding the pancaking effect on the structure formation.  

In the light of the first detection of NL06, a couple of questions naturally 
arise: When does the pancaking effect influence most dominantly the formation 
of dark halos?; How does the pancake scale change with time?; What are the 
implications and consequences of the pancaking effect on the structure 
formation?; Is the standard theory capable of quantifying the velocity 
anisotropy?. Our goal here is to provide answers to these questions.

We organize this paper as follows. In \S 2, we introduce the analyzing 
method to investigate the evolution of the pancaking effect in the Millennium
Simulation. We describe analytic modeling and compare it with the results 
obtained in the simulation catalog in \S 3. Finally, we discuss and conclude 
the results.

\section{SIGNALS FROM SIMULATION}

To investigate the evolution of local pancaking effect, we use the Millennium 
Run simulation halo catalogs for the concordance $\Lambda$CDM cosmology 
\citep{spr-etal05}. A periodic box of this simulation is $500\ h^{-1}$Mpc on 
each side and the cosmological parameters are $\Omega_{m}=0.25$, 
$\Omega_{\Lambda}=1-\Omega_{m}$, $h=0.73$, and $\sigma_{8}=0.9$. We choose four 
different redshift catalogs, approximately $z=0,\ 0.5,\ 1$, and $2$ among 64 
output times of the simulation. Then we analyze the alignments of dark-matter halo 
velocities to the local pancake plane in each catalog. Basically, to find a local
pancake, we follow the methodology suggested by NL06.

As a first step in measurement, we select the halos which contain more than 
$50$ particles (the left column of Table \ref{tab:num}) since those halos in 
Millennium Run catalog which have too low particle numbers (less than $30$) 
are poorly sampled and severely contaminated by the numerical noise. 
In other words, only those halos which have enough particles to define a halo 
density profile are reliable (V. Springel in private communication). 
Thus, we set the particle number cut-off at 50. Next, among the selected halos, 
we pick out field halos which are still in quasi-linear regime by using the 
method suggested by \citet{ela-pir97} and \citet{hoy-vog02}. The number of 
identified field halos is shown in the right column of Table \ref{tab:num}. 
The field halos are suited for our purpose because they may not be seriously 
affected by gravitational forces of many other halos, retaining their 
initial conditions at the moment that they collapsed.

Using only those field halos, we determine local pancakes using the 
practical methods proposed by NL06. First, We find two nearest field 
halos for a field halo and define a local pancake plane enclosing all the 
three halos. Here, the local pancakes are found by changing the criterion 
distance $R_{c}$ from $0$ $h^{-1}$Mpc to $17$ $h^{-1}$Mpc which is a lower 
limit of $R_{1}$, the displacement vector to the first nearest halo. 
In other words, $R_{c}$ satisfies the following condition: 
$R_{c} \le R_{1} \le R_{2}$ where $R_{2}$ is the displacement vector to the 
second nearest neighbor.

With the determined local pancakes, then we calculate the alignment angles between 
a halo velocity vector and a vector normal to the local pancake plane, $\cos\theta$.
Finally, we obtain the probability density distribution of $\cos\theta$, 
$p(\cos\theta)$, by counting the number of halos in each bin of $\cos\theta$.
We note that in order to examine the dependence of the particle number cut-off 
which is used when we determine the halos, we recalculate the velocity-pancake
alignment for those halos which include more than $30$ particles though 
low-particle number halos suffer from simulation noise. From this calculation, 
we obtain almost perfectly consistent results with the results of the catalog 
in which halos include more than $50$ particles as can be seen in 
Fig. \ref{fig:pro} (particle number cut-off is $50$) and Fig. \ref{fig:pro30} 
(particle number cut-off is $30$). Thus, the velocity-pancake alignment is 
not affected by the choice of particle number cut-off.

Figure \ref{fig:pro} shows $p(\cos\theta)$ with Poissonian error. The 
three columns correspond to three different cases of $z=0,\ 1$, and $2$, 
respectively, and the three rows to three different scales of $R_{c}=0,\ 3$, 
and $6$, respectively. The horizontal dotted line in each panel represents a random 
distribution. As can be seen in the right column, when $R_{c}$ is $0$, the halo 
velocity tends to be slightly aligned with the normal to the local pancake plane 
at $z=2$. This tendency becomes stronger as $R_{c}$ increases to $6 h^{-1}$Mpc. 
It can be interpreted that the local pancaking effect arises at least around 
$z\simeq 2$. Meanwhile, at $z=1$, for the case of $R_{c}=0$, there is no alignment. 
Then local pancaking effect starts to be visible near $R_{c}=3 h^{-1}$Mpc. 
Moreover, as one can see in the left column, when $R_{c}=0$ at $z=0$, the halo 
velocity does not tend to be aligned with the normal to the local pancake plane 
any more. Rather it tends to lie on the plane. The alignment signal begins to be 
seen when $R_{c}$ is approximately $6 h^{-1}$Mpc. We infer from these phenomena 
that the pancaking effect disappears since the gravitational attraction forces 
from the close neighbor halos affect halo velocities rather than the initial 
pancaking effect. Therefore, we find the velocity-pancake alignments are 
strongest at $z\simeq2$, indicating that the pancaking effect starts at 
$z\geqq 2$. We refer that we do not make more analysis of the velocity-pancake 
alignments for higher redshift to find a redshift at which pancaking effect 
starts since the number of halos decreases rapidly as redshift increases. 
Even if the value of $\eta$ itself might increase, the value of $\sigma_{\eta}$ 
which is inversely proportional to the number of halos would increase.
Thus, it would be hard to determine precisely at which redshift both the 
values of $\eta$ and $\sigma_{\eta}$ reach maximums. 

Figure \ref{fig:his} plots the average of $\cos\theta$ at four different $z=0,\ 
0.5,\ 1$, and $2$ as a function of linear characteristic pancake scale, $L_{p}$.
We refer that $L_{p}$ is defined as the average distance between the center of 
mass (CM) for three halos on the local pancake plane to all the three halos, 
which was proposed by NL06. The horizontal dotted line in each panel represents 
$\langle\cos\theta\rangle$ of random distribution of the angles. As one can see
in each panel, there is a specific range of $L_{p}$ where $\langle\cos\theta\rangle$ 
is relatively higher than those at the other pancake scales. This specific range 
implies local pancaking effect significantly occurs on specific pancake scale. 
Comparing to panels each other, overall $\langle\cos\theta\rangle$ increases, its 
slope is steeper and the specific range shifts to small $L_{p}$ when going to high 
redshift. This tendency can be clearly seen when we see that the regions of filled with 
oblique lines, which represent the area consisting of a few of the highest values, 
move to small $L_{p}$ as going to high redshift. More quantitatively, the peak value 
is approximately $5.035\times 10^{-4}$ at $z=0$ while the one is $5.055\times 
10^{-4}$ at $z=2$. In addition, the peak position of histogram is approximately 
$2\ h^{-1}$Mpc at $z=0$ whereas the one is $8\ h^{-1}$Mpc at $z=2$. 
This gradual change with redshifts can be understood the same way as
Fig. \ref{fig:pro}; after pancaking effect on halo velocities arises, 
it is gradually attenuated by gravitational force among halos, especially among 
very close halos. 

\section{THEORETICAL ANALYSIS}
In order to investigate the tendency of the anisotropic distribution of halo 
velocity theoretically, we construct an analytic model using the similar 
methodology to \citet{lee-pen01}. We begin on the Zel'dovich 
approximation \citep{zel70}
\begin{equation}
\label{eq:zel}
{\bf x}={\bf q}-D(t)\nabla\Psi({\bf q}), 
\end{equation}
which intrinsically estimates the formation of pancakes. Here, ${\bf x}$ is the 
Eulerian coordinate, ${\bf q}$ is the Lagrangian coordinate, $\Psi({\bf q})$ is
the linear velocity potential and $D(t)$ is the growth factor,
\begin{equation}
\label{eq:gro}
D=\frac{5}{2}\Omega_{m}H(a)\int_{0}^{a}\frac{da'}{[a'H(a')]^{3}}.
\end{equation}
By taking the time derivative of Zel'dovich approximation, we obtain equation
of the velocity of cosmic particle smoothed on pancake scale,
\begin{equation}
\label{eq:xdot}
v_{i}=-({\dot D}\partial_{i}\Psi + D\partial_{i}{\dot \Psi}) \propto
-({\dot D}q_{j}T_{ij}+Dq_{j}{\dot T}_{ij}).
\end{equation}
Equation (\ref{eq:xdot}) is appropriate to investigate pancaking effect on halo 
velocity at early stage since equation (\ref{eq:zel}) describes initial state of 
non-linear evolution. Also, we assume the density field is smoothed on the pancake 
scale, which is valid in this approximation because this approximation is available 
in the regime preceding the moment of the collapse to a pancake. 

Using a similar logic given by \citet{lee-pen01} who found an expression for the 
alignments between the tidal field and the halo position vectors, we take into 
account nonlinear effects into equation (\ref{eq:xdot}), and write an expression 
for the velocity alignments with the tidal field as
{\setlength\arraycolsep{2pt}
\begin{eqnarray}
\label{eq:velcor}
\langle v_{i}v_{j}\vert {\bf T}\rangle&=&\frac{1-\eta}{3}\delta_{ij}+\eta K_{ij},
\\ 
\label{eq:K} K_{ij}&\equiv& {\dot D}^{2}T_{ik}T_{kj} +
2D{\dot D} T_{ik}{\dot T_{kj}} + D^{2}{\dot T}_{ik}{\dot T}_{kj}
\end{eqnarray}}
where $\bf T$ and $\dot{\bf T}$ are tidal shear tensor and its time derivative,
respectively. $\dot D$ is time derivative of $D$, which is
{\setlength\arraycolsep{2pt}
\begin{eqnarray}
\label{eq:ddot}
& &{\dot D}=\frac{5}{2}\Omega_{m}\left[H'(a){\dot a} \int_{0}^{a}\frac{da'}
{\{a'H(a')\}^{3}}+ \frac{{\dot a}H(a)}{a^{3}H(a)}\right],\\
\label{eq:dhda}
\textrm{where}\quad H'(a)&=&\frac{dH(a)}{da}=-\frac{3}{2}\Omega_{m}a^{-4}(\Omega_{m}
a^{-3}+\Omega_{\Lambda})^{-1/2}, \quad
{\dot a}=aH_{0}(\Omega_{m}a^{-3}+\Omega_{\Lambda})^{1/2}.
\end{eqnarray}}
Notice that $\eta$ is a value in $[0,1]$. If $\eta$ is $1$, it represents
a perfect correlation between local pancaking and halo velocity while if $\eta$ 
is $0$, it means there is no correlation between them. In velocity-velocity 
correlation, it is unlikely that the value of $\eta$ is close to unity
since the velocity of a halo is not a conserved quantity and initial memory of 
local pancaking effect on velocity have been reduced. 
In order to show that the velocities of halos are not completely random but 
has some degree of alignments with the normal vectors to the local pancakes
though The value of $\eta$ is expected to be very small, we have calculated 
the error on the value of $\eta$, that is, the standard deviation of $\eta$ for 
the case of no alignment. The formula for the error of $\eta$, $\sigma_{\eta}$ 
is given as $(4/15N_{tot})^{1/2}$ where $N_{tot}$ is the total number of halos 
\citep{lee-pen01}. With defined error, if $\eta$ is larger than $3\sigma_{\eta}$, 
one can say that there is a signal of true local pancaking effect.

We diagonalize equation (\ref{eq:velcor}) by applying the relation referred by
\citet{bon-etal96} 
\begin{equation}
\label{eq:bond}
{\dot\lambda_{i}}=\frac{D\lambda_{i}}{1-D\lambda_{i}} 
\end{equation}
where $\lambda_{i}$ and ${\dot \lambda_{i}}$ are an eigenvalue of the ${\bf T}$ 
and its time derivative, respectively. Thus, equation (\ref{eq:K}) 
is changed to 
\begin{equation}
\label{eq:kdiag}
K_{ii}=\frac{\lambda_{i}^{2}[{\dot D}(1-D\lambda_{i})+D^{2}]^{2}}
{(1-D\lambda_{i})^{2}},\quad
K_{ij}=0 \; (i\ne j).
\end{equation}
The eigenvalues $\lambda_{i}$ are ordered by $\lambda_{1}>\lambda_{2}>\lambda_{3}$.

To obtain the probability distribution, we assume probability density distribution of 
the alignment angles, $\cos\theta$, is Gaussian \citep{lee04} 
\begin{equation}
\label{eq:pden}
p({\hat v}\vert{\check {\bf T}})=\frac{1}{\sqrt{(2\pi)^{3}\det(M)}}\int_{0}^
{\infty}\exp\left[-\frac{v_{i}^{T}(M^{-1})_{ij}v_{j}}{2}\right]v^{2}dv,
\end{equation}
where $M_{ij}\equiv\langle v_{i}v_{j}\vert {\bf T}\rangle$.
Finally, we can express the probability distribution of $\cos\theta$, $p(\cos
\theta)$ is,
\begin{equation}
\label{eq:prob}
p(\cos\theta)=\frac{1}{2\pi}\prod_{i=1}^{3}(1+\eta-3\eta\check{\lambda}_{Ki})^
{-\frac{1}{2}}\int_{0}^{2\pi}\left(\frac{\sin^{2}\theta\cos^{2}\phi}{1+\eta-3\eta
\check{\lambda}_{K1}}+\frac{\sin^{2}\theta\sin^{2}\phi}{1+\eta-3\eta
\check{\lambda}_{K2}}+\frac{\cos^{2}\theta}{1+\eta-3\eta\check{\lambda}_{K3}}
\right)^{-\frac{3}{2}}d\phi.
\end{equation}
Here, ${\check \lambda}_{Ki} = {K_{ii}}/{\sum_{i} K_{ii}^{2}}$.

To compare with the results from simulation analysis, we use $\Lambda$CDM 
cosmological parameters, $h=0.73$, $\sigma_{8}=0.9$, $T_{cmb,0}=2.725$K,
$\Omega_{b}=0.045$, $\Omega_{m}=0.25$, which are the same values as 
the parameters used in Millennium Simulation. 
First, we calculate $\sigma$ by using following equation: 
\begin{equation}
\label{eq:sig}
\sigma^{2} (R,z)=D(z)^{2}\int \frac{k^{2}dk}{2\pi ^{2}}P(k)T^{2}(k)
W_{TH}^{2}(kR),
\end{equation}
where $D(z)$ is a growth function, $P(k)$ is primordial power spectrum that is $P(k)
\propto k$, $T(k)$ is transfer function \citep{bar-etal86} and $W_{TH}$ is spherical 
top-hat window function.
When evaluating equation  (\ref{eq:sig}), we use $L_{p}$ obtained by simulation 
analysis for $R$. Also, the redshift at which $L_{p}$ is determined in the simulation
is chosen as $z$ value in equation (\ref{eq:sig}).

Then we obtain $\lambda_{Ki}$ in equation (\ref{eq:pden}). We decide 
$\lambda_{K1}$ as $1$ because $\lambda_{1}=1$ means the formation of the structure
like a pancake in the Zel'dovich approximation,
\begin{equation}
\label{eq:rho} 
\rho = \frac{\bar{\rho}}{(1-\lambda_{1})(1-\lambda_{2})(1-\lambda_{3})},
\end{equation}
where $\bar{\rho}$ is the mean density of the universe and $\lambda_{1},\ \lambda_{2},\ 
\lambda_{3}$ ($\lambda_{1} > \lambda_{2} > \lambda_{3}$) are the eigenvalues of the 
local tidal field, ${\bf T}$. 
For $\lambda_{K2}$ and $\lambda_{K3}$, we choose the most probable values of the 
following probability distribution, 
\begin{equation}
\label{eq:dor}
p(\lambda_{2},\lambda_{3}\vert\lambda_{1}=1) = p(\lambda_{1}=1,\lambda_{2},
\lambda_{3})/p(\lambda_{1}=1).
\end{equation}
Here, $p(\lambda_{1},\lambda_{2},\lambda_{3})$ and $p(\lambda_{1})$ are
the probability distribution derived by \citet{dor70} and \citet{lee-sha98}:
\setlength\arraycolsep{1pt}
\begin{eqnarray}
\label{eq:pl1l2l3}
&p&(\lambda_{1},\ \lambda_{2},\ \lambda_{3})=
\frac{3375}{8\sqrt{5}\pi\sigma^{6}}(\lambda_{1}-\lambda_{2})(\lambda_{2}
-\lambda_{3})(\lambda_{1}-\lambda_{3})\exp\left (-\frac{3I_{1}^{2}}{\sigma^{2}}
+\frac{15I_{2}}{2\sigma^{2}}\right )\\
\label{eq:pl1}
&p&(\lambda_{1})=
\frac{\sqrt{5}}{12\pi\sigma}\Bigg[
20\frac{\lambda_{1}}{\sigma}\exp\left(-\frac{9\lambda_{1}^{2}}
 {2\sigma^{2}}\right)-\sqrt{2\pi}\exp\left(-\frac{5\lambda_{1}^{2}}{2\sigma^{2}}\right)
 \textrm{erf}\left(\sqrt{2}\frac{\lambda_{1}}{\sigma}\right)
 \left(1-20\frac{\lambda_{1}^{2}}{\sigma^{2}}\right) \nonumber \\
& &\quad\quad -\ \sqrt{2\pi}\exp\left(-\frac{5\lambda_{1}^{2}}
  {2\sigma^{2}}\right) \left(1-20\frac{\lambda_{1}^{2}}{\sigma^{2}}\right) 
+3\sqrt{3\pi}\exp\left(-\frac{15\lambda_{1}^{2}}{4\sigma^{2}}\right)
\textrm{erf}\left(\frac{\sqrt{3}\lambda_{1}}{2\sigma}\right) \nonumber \\
& &\quad\quad +\ 3\sqrt{3\pi}\exp\left(-\frac{15\lambda_{1}^{2}}{4\sigma^{2}}\right)\Bigg].
\end{eqnarray}
We put these eigenvalues in equation (\ref{eq:kdiag}) to obtain the normalized eigenvalues,
$\check{\lambda}_{Ki}$.

\begin{figure} \begin{center} \epsscale{1.0} \plotone{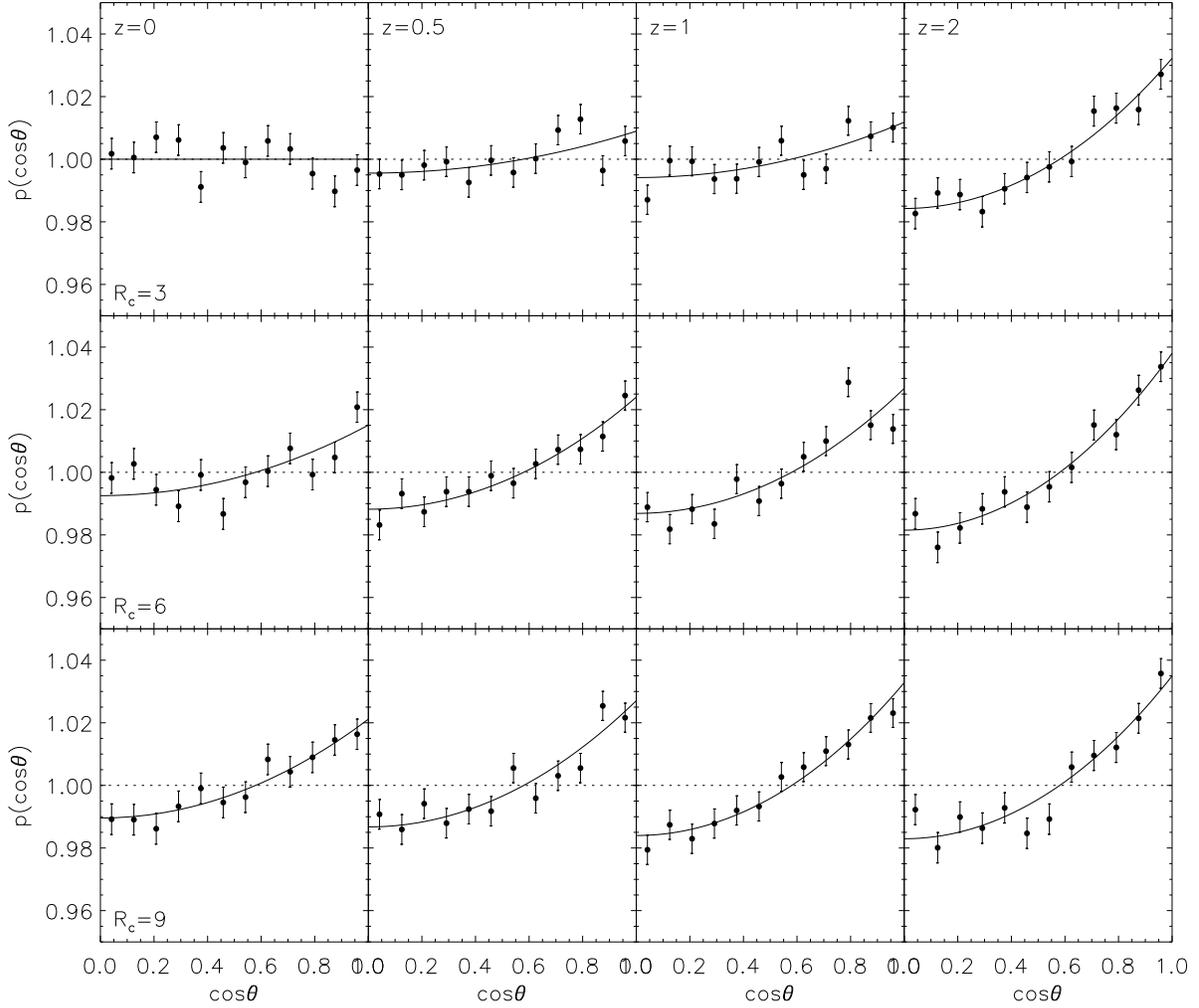}
\caption{Probability density distributions of the $\cos\theta$s. Filed circle with
Poissonian errors represent the results from the simulation while solid line stand for
the analytic prediction with correlation parameter $\eta$. The horizontal dotted lines
correspond to no correlation.}
\label{fig:the}
\end{center}
\end{figure}

Finally, we fit theoretical estimation of $p(\cos\theta)$, equation
(\ref{eq:prob}), to the numerical results as adjusting
a correlation parameter $\eta$ in equation (\ref{eq:prob}).
We determine $\eta$ minimizing $\chi^{2}$ distribution
\begin{equation}
\label{eq:chi}
\chi^{2}=\sum_{i}\frac{x_{s,i}-x_{t,i}}{\sigma_{s,i}},
\end{equation}
where $x_{s,i}$ and $x_{t,i}$ are the values from simulation analysis and from theory, 
respectively and $\sigma_{s,i}$ is the Poissonian error from simulation analysis. 
The comparison numerical results with theory is shown in Fig. \ref{fig:the}.
The filled circle points represent the numerically measured $p(\cos\theta)$ with 
Poissonian errors and solid line represents the analytic distribution. As can be seen
in the figure, the simulation data points are in good agreement with our analytic 
formulation. Also, as shown in Table \ref{tab:eta}, the strength of local pancaking 
effect on the halo velocity depends on the redshift. At $z=2$, though halos are 
closely located at each other like $R_{c}=3 h^{-1}$Mpc, $\eta$ is 
$0.011$, which turns out to be $15.3\sigma_{\eta}$ (see Table \ref{tab:eta}). 
Thus, even though the value of $\eta$ is much less than unity, it is 
definitely a strong signal of velocity-pancake alignments, exceeding $15$ times 
the standard deviation, $\sigma_{\eta}$. This strong signal which approximately 
amounts to $15\sigma_{\eta}$ shows the existence of the velocity-pancake alignment.
Even when $R_{c}=0$ at $z=2$, $\eta$ is $0.003$ which is $4\sigma_{\eta}$ 
which indicates that a pancaking effect appears even at $R_{c}=0$. Then going to present epoch, the 
effect is reduced by gravitational attraction. Finally, at $z=0$, when $R_{c}$ is 
$3 h^{-1}$Mpc, $\eta$ is almost $0$ and $\sigma_{\eta}$ is also $0$. In addition, 
as one can see the tendency in Fig. \ref{fig:the} and in Table \ref{tab:eta}, at 
$z=2$, pancaking effect becomes stronger as $R_{c}$ is approximately $5-6 h^{-1}$Mpc 
and then $\eta$ and $\sigma_{\eta}$ gradually decreases when $R_{c}$ is larger than 
$6 h^{-1}$Mpc. In other words, the value of $\eta$ and $\sigma_{\eta}$ at $z=2$ is 
the largest with $R_{c}\simeq 6$. A similar tendency also appears at different redshifts, 
instead the range of $R_{c}$ is shifted to larger scale caused by the expansion of 
the universe. Thus, we analytically find that there is a specific range of pancake 
scale where pancaking significantly affects halo velocity. In addition, we show how 
the pancaking effect changes with different redshifts. 

\begin{figure} \begin{center} \epsscale{1.0} \plotone{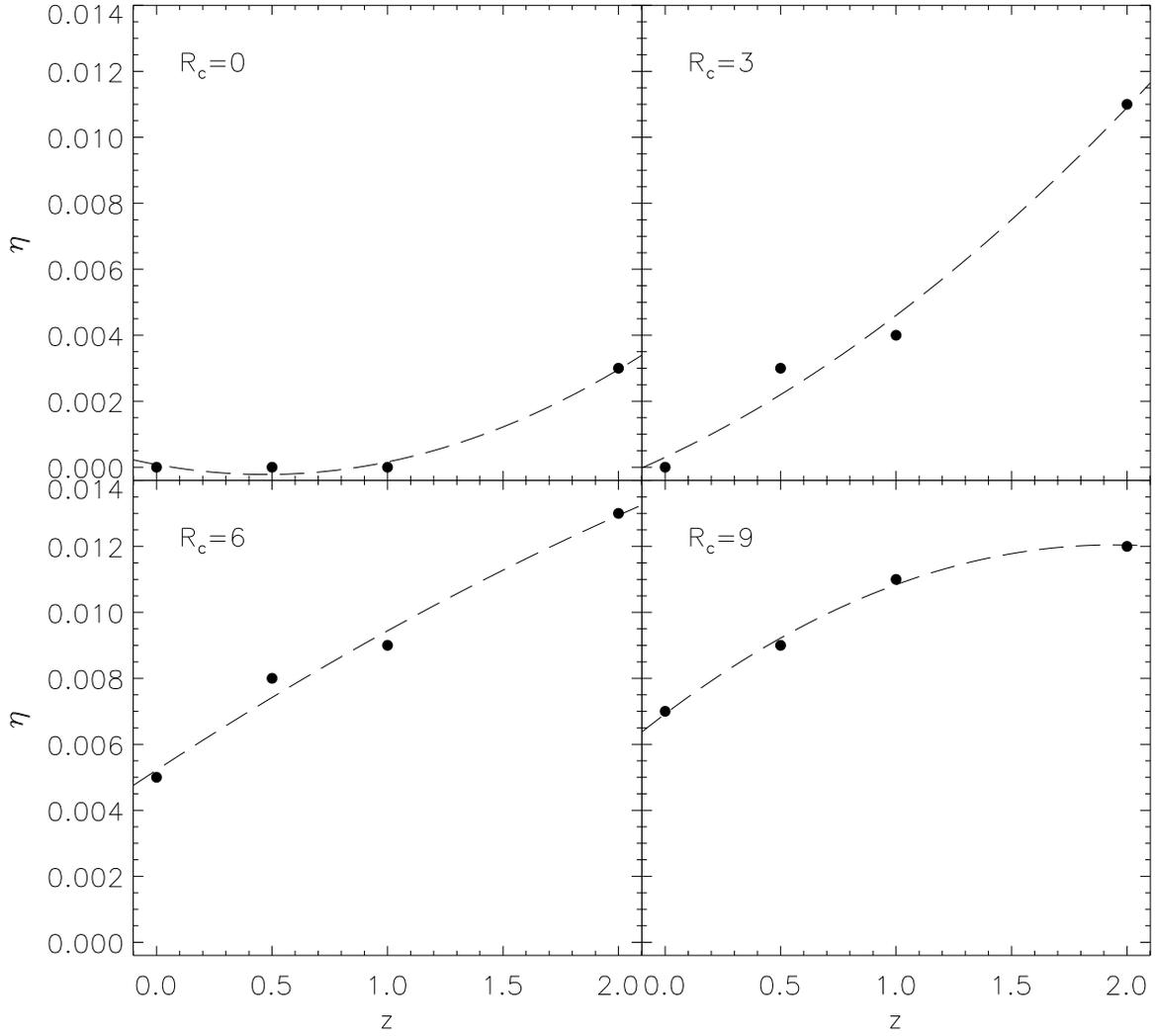}
\caption{$\eta$ as a function of redshifts. Filed circle represents the numerical 
measurement in Millennium Simulation. Dashed line is a quadratic fitting function 
at different $R_{c}$ scale.}
\label{fig:zco}
\end{center}
\end{figure}

\begin{figure} \begin{center} \epsscale{1.0} \plotone{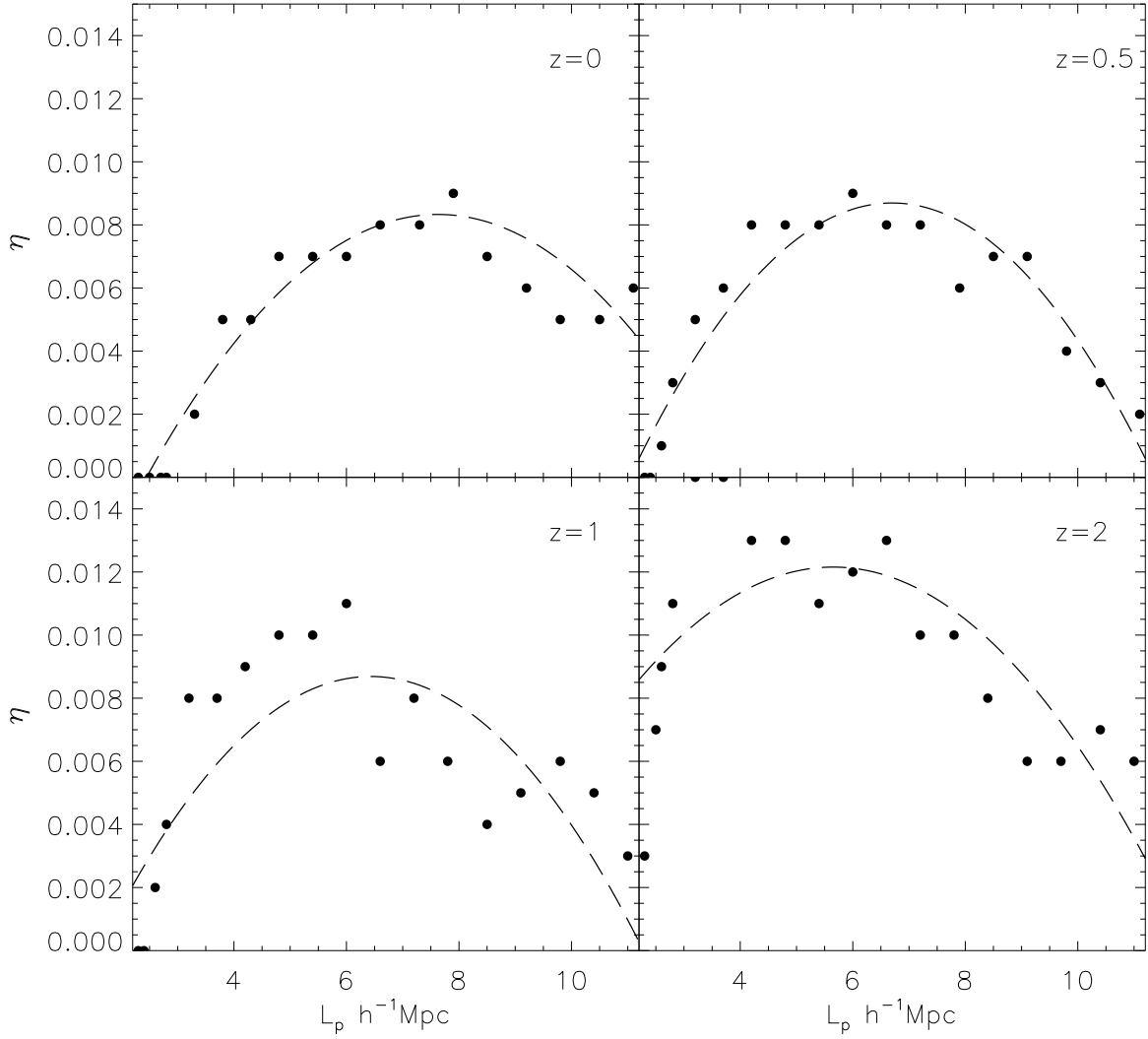}
\caption{The same as Fig. \ref{fig:zco} but as a function of characteristic pancake 
scale, $L_{p}$.}
\label{fig:rco}
\end{center}
\end{figure}

To find the tendency of the redshift dependence of $\eta$, we fit $\eta$ as a function 
of redshift to quadratic equation. The coefficients are shown in Table \ref{tab:fz}. 
Figure \ref{fig:zco} plots the quadratic fitting function (dashed line) and compares
it with the simulation data points (filled circles). As Fig. \ref{fig:zco} shows,
we note that $\eta$ is a function of redshift, in a good agreement with the quadratic 
equation. Moreover, $\eta$ is a function of the characteristic pancake scale, 
$L_{p}$. We also fit $\eta(L_{p})$ to quadratic equation which can be seen in 
Fig. \ref{fig:rco}. The dashed line and the filled circles represent the fitting 
function and the data points, respectively. We can see a good match between the 
quadratic functional form and the simulation data. The best-fit coefficients are shown 
in Table \ref{tab:fLp}.

\section{DISCUSSION AND CONCLUSION}

We analyze the evolution of the pancaking effect by using millennium simulation.
We obtain the signal of the strong alignment between the halo velocity and the 
normal to the local pancake plane at $z\simeq 2$ when characteristic pancake 
scale is approximately $3\ h^{-1}$Mpc. The alignment signal implies the first
collapse of the protocloud may have occurred around $z\simeq 2$ and the size 
of the pancake may have been around $3\ h^{-1}$Mpc.
This result is in agreement with \citet{mo-etal05} which 
reports pancakes form at $z\simeq 2$ and their masses is approximately $5\times 
10^{12} \textrm{M}_{\odot}$. 

\begin{figure} \begin{center} \epsscale{1.0} \plotone{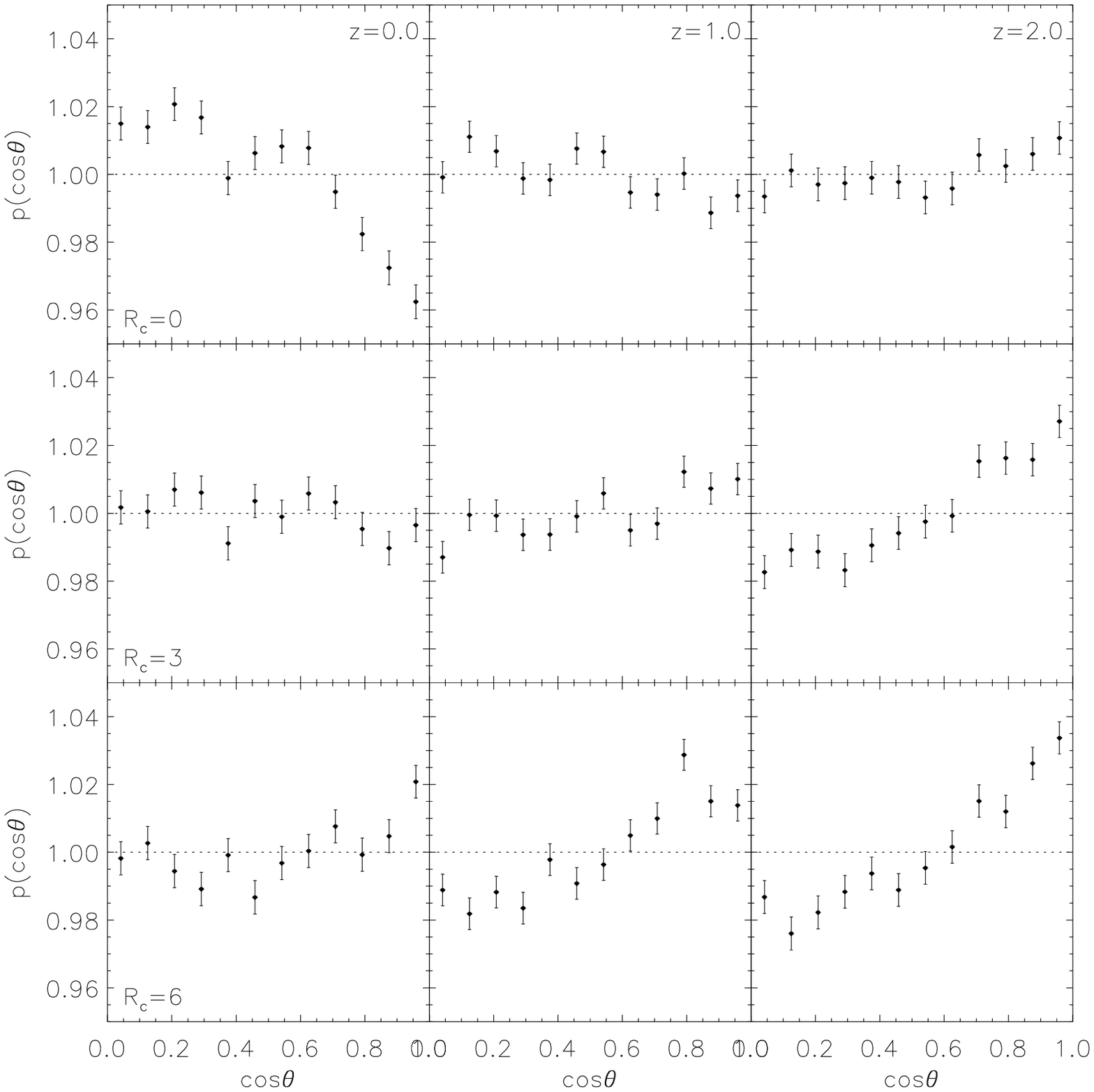}
\caption{Probability density distributions of the cosines of the relative 
angles between the field halo velocities and the directions 
normal to the local planes enclosing the two neighbor field halos 
for the cases of the distance threshold: $R_{c}=0,\ 3,\ 6\ h^{-1}$Mpc 
at different redshift: $z=0,\ 1,\ 2$ from Millennium Run simulation.
The errors are Poissonian and 
the horizontal dotted lines represent no alignment. Note that halos which
contain more than $50$ particles are determined} 
\label{fig:pro}
\end{center}
\end{figure}

\begin{figure} \begin{center} \epsscale{1.0} \plotone{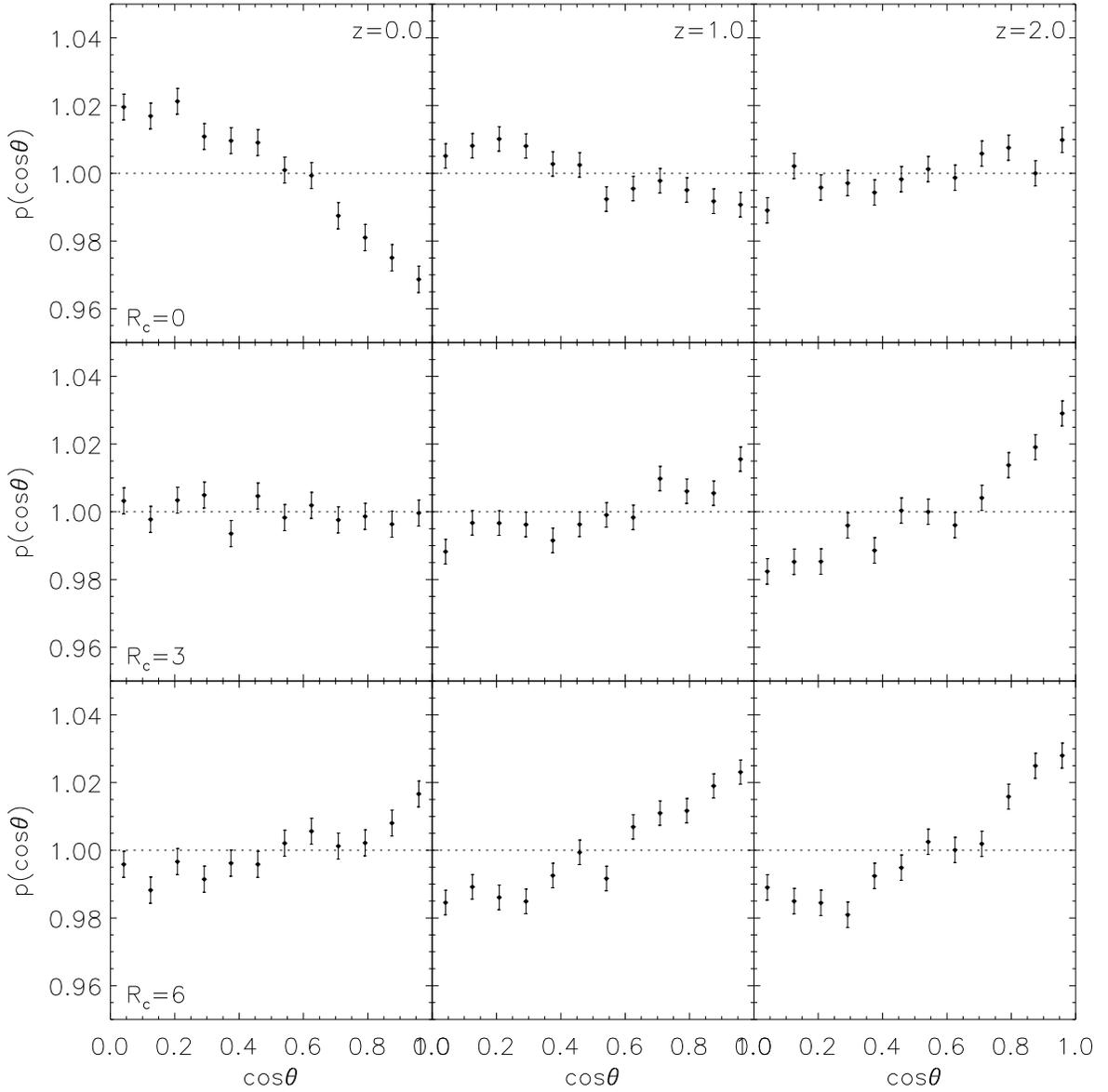}
\caption{The same figure as Fig. \ref{fig:pro} but using the halos which 
contain more than $30$ particles.}
\label{fig:pro30}
\end{center}
\end{figure}

\begin{figure} \begin{center} \epsscale{1.0} \plotone{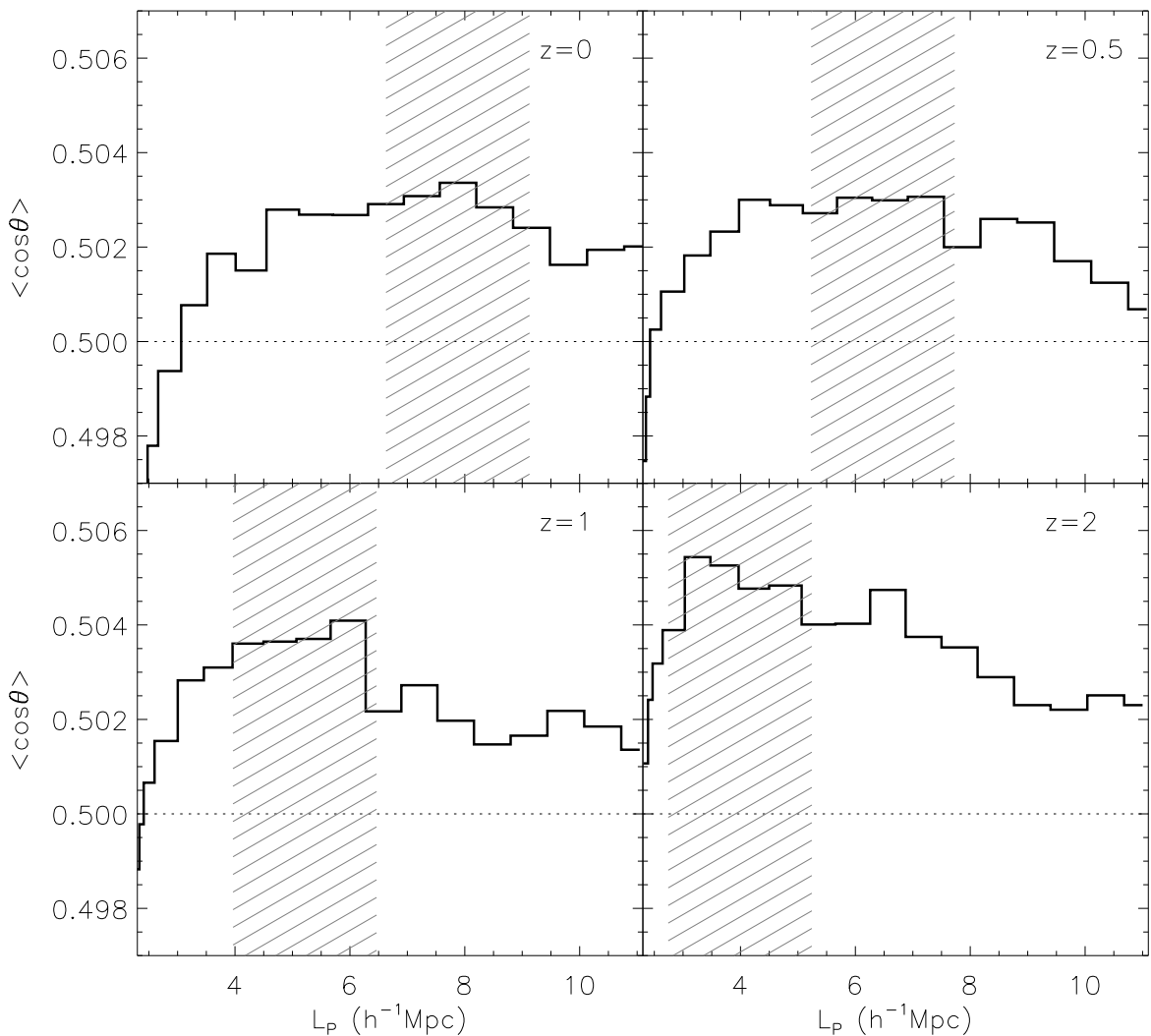}
\caption{The average of the cosines angles ($\cos\theta$) between the halo 
velocities and the local pancake planes whose definition is the
same as Fig. \ref{fig:pro}, corresponding to the change of characteristic 
linear size of the pancake, $L_{p}$ at four different redshift: $z=0,\ 0.5,\ 1,\ 2$.
The horizontal dotted lines represent no correlation.}
\label{fig:his}
\end{center}
\end{figure}

Then as shown in Fig. \ref{fig:pro} and Fig. \ref{fig:his}, the alignment signal 
evolves. Pancaking effect is weakened as going to low redshift, which means the 
pancaking effect on halo velocity is gradually attenuated by the gravitational 
attraction among the neighbor halos. In addition, pancake scale at low redshift 
is larger than that at high redshift.

We expect the local pancaking effect could be detected in observation.
It may be possible to achieve the signal of the velocity alignment using the velocity 
data reconstructed from 2MASS redshift survey \citep{erd-etal06}. 
We intend to measure the signal in observation for our future project.

It is interesting to note that the pancakes are expected to have formed coincidentally
at $z\simeq 2$ when the star formation rate in massive galaxies is the highest 
\citep{jun-etal05,feu-etal05}, which is one of the phenomena referred to `downsizing'.
In `downsizing' scenario, the stars in more massive galaxies formed at higher redshift 
and those in less massive galaxies recently formed, where `antihierarchy' can be 
suggested. Recently, however, \citet{nei-etal06,mou-tan06} show this phenomenon is 
natural in hierarchical clustering process. Nevertheless, a coincidence between the 
pancake formation epoch and the epoch of the heavily star-forming may imply the 
structure formation is not simply hierarchical but complicated \citep{cim-etal06}.

In conclusion, we obtain the result that is the evolution of the velocity anisotropy 
induced by local pancake formation. This result would help to understand the unknowns
in the galaxy formation.

\acknowledgments 
The Millennium Run simulation used in this paper was carried out 
by the Virgo Supercomputing Consortium at the Computing Centre of the 
Max-Planck Society in Garching (http://www.map-garching.mpg.de/millennium). 
We are thankful to L. Gerard for the halo catalogs 
of the Millennium Run simulation and to V. Springel for useful discussion.
We also appreciate the anonymous referee for helpful suggestions.
This work is supported by the research grant No. R01-2005-000-10610-0 from 
the Basic Research Program of the Korea Science and Engineering Foundation.

\clearpage
\clearpage
\clearpage
\clearpage
\clearpage
\clearpage
\clearpage
\begin{deluxetable}{ccc}
\tablewidth{0pt}
\setlength{\tabcolsep}{5mm}
\tablehead{\colhead{$z$}&\colhead{$N_{total}$}&\colhead{$N_{field}$}
($\case{N_{field}}{N_{total}}$ \%)}
\tablecaption{{\it left}: Total number of identified halos whose contain
more than $50$ particles in each redshift catalog, {\it right}: The number of 
field halos \label{tab:num}}
\startdata
$0$ & $6126689$ & $502108$ ($8.2$) \\
$0.5$ &$6631145$ & $542554$ ($8.2$)\\
$1$ & $6817225$ & $558138$ ($8.2$)\\
$2$ & $6222187$ & $516774$ ($8.3$)\\
\enddata
\end{deluxetable}
\clearpage
\begin{deluxetable}{ccccc}
\tablewidth{0pt}
\setlength{\tabcolsep}{5mm}
\tablecaption{Correlation parameter $\eta$ corresponding to $R_{c}$ and
redshift, $z$ with its standard deviation for the case of no alignment. 
The formula for the error of $\eta$, $\sigma_{\eta}$ is given as $(4/15N_{tot})^{1/2}$ 
where $N_{tot}$ is the total number of halos \citep{lee-pen01}. \label{tab:eta}}
\tablehead{\colhead{$R_{c}$} &\multicolumn{4}{c}{z}\\
\cline{2-5}
\colhead{($h^{-1}$Mpc)} & \colhead{$0$} & \colhead{$0.5$}   & \colhead{$1$}     & \colhead{$2$}    }
\startdata
$3$ & $0.000$($0.0\sigma_{\eta}$) & $0.003$($4.3\sigma_{\eta}$) & 
$0.004$($5.8\sigma_{\eta}$) & $0.011$($15.3\sigma_{\eta}$) \\
$6$ & $0.005$($6.9\sigma_{\eta}$) & $0.008$($11.4\sigma_{\eta}$) & 
$0.009$($13.0\sigma_{\eta}$) & $0.013$($18.1\sigma_{\eta}$) \\
$9$ & $0.007$($9.6\sigma_{\eta}$) & $0.009$($12.8\sigma_{\eta}$) & 
$0.011$($15.9\sigma_{\eta}$) & $0.012$($16.7\sigma_{\eta}$) \\
\enddata
\end{deluxetable}
\clearpage
\begin{deluxetable}{cccc}
\tablewidth{0pt}
\setlength{\tabcolsep}{5mm}
\tablehead{\colhead{$z$}&\colhead{$a$}&\colhead{$b$}&
\colhead{$c$}}
\tablecaption{Fitting parameters for $\eta$ as a function of redshift, $z$.
Fitting functions are quadratic function, $\eta (z)=az^{2}+bz+c$, ($10^{-3}$)
\label{tab:fz}}
\startdata
$0$    & $1.36$\phd & $-1.28$ & $0.08$ \\
$0.5$  & $1.00$\phd & $3.30$ & $0.30$ \\    
$1$    & $-0.36$\phd & $4.58$ & $5.22$ \\    
$2$    & $-1.36$\phd & $5.28$ & $ 6.92$ \\    
\enddata
\end{deluxetable}
\clearpage
\begin{deluxetable}{cccc}
\tablewidth{0pt}
\setlength{\tabcolsep}{5mm}
\tablehead{\colhead{$z$}&\colhead{$a$}&\colhead{$b$}&
\colhead{$c$}}
\tablecaption{The same as Table \ref{tab:fz} but as a function 
of linear characteristic pancake scale, $L_{p}$, $\eta (L_{p})=aL_{p}^{2}+bL_{p}
+c$, ($10^{-3}$) \label{tab:fLp}}
\startdata
$0$   & $-0.31$ & $4.73$ & $-9.71$\\
$0.5$ & $-0.40$ & $5.36$ & $-9.26$\\   
$1$   & $-0.37$ & $4.76$ & $-6.62$\\
$2$   & $-0.30$ & $3.39$ & $2.58$\\
\enddata
\end{deluxetable}

\end{document}